\documentclass[epj]{svjour}
\usepackage{amsmath,amssymb,amsfonts,graphicx,cite,algorithmic,algorithm,multirow}
\graphicspath{{graphics/}}
\makeatletter
\ifx\input@path\@undefined
\def\input@path{{graphics/}}
\else
\g@addto@macro\input@path{{graphics/}}
\fi
\makeatother
\newcommand{\program}[1]{\textsf{#1}}

\newenvironment{inlinealgorithm}{%
\vspace*{2ex}
\hrule
\begin{small}
\begin{algorithmic}
}{%
\end{algorithmic}
\end{small}
\hrule
\vspace*{2ex}
}

\preprint{DESY 11-146\\KA-TP-22-2011\\MCnet-11-19}

\title{The Sudakov Veto Algorithm Reloaded}

\author{Simon Pl\"atzer\inst{1} \and Malin Sj\"odahl\inst{2}}

\institute{DESY, Notkestrasse 85, D-22607 Hamburg, Germany\and%
Institut f\"ur Theoretische Physik, KIT, D-76128 Karlsruhe, Germany}

\date{\today}

\abstract{ We perform a careful analysis of the main Monte Carlo
  algorithm used in parton shower simulations, the Sudakov veto
  algorithm. We prove a general version of the algorithm, directly
  including the dependence on the infrared cutoff. Taking this as a
  starting point, we then consider non-positive definite splitting
  kernels, as encountered when dealing with sub-leading colour
  correlations or splitting kernels beyond leading order.  New
  algorithms suited for these situations are developed.   }

\begin{document}

\maketitle


\section{Introduction}

Parton shower Monte Carlo simulations as implemented in for example
\cite{Bahr:2008pv,Sjostrand:2007gs,Gleisberg:2008ta}, 
are indispensable tools for analyzing
and predicting realistic final states encountered in collider
experiments. Matrix element corrections, as discussed in
\cite{Seymour:1994df,Bengtsson:1986hr,Bengtsson:1986et,Norrbin:2000uu,Miu:1998ju},
the technically similar matching to NLO calculations employing the
POWHEG method \cite{Nason:2004rx}, or schemes to combine parton
showers and multijet tree-level matrix elements
\cite{Catani:2001cc,Lonnblad:2001iq,Giele:2007di,Hoeche:2009rj,Hamilton:2009ne},
all rely directly or indirectly on the same method for generating
subsequent parton shower emissions in Monte Carlo simulations. 

With the notable exception of the FORTRAN version of \program{HERWIG},
nowadays most parton shower implementations use the Sudakov veto
algorithm to facilitate this task, as the splitting kernels normally 
are too complicated to allow efficient integration.

A justification of the Sudakov veto algorithm is given in
\cite{Seymour:1994df}, stating that
for upper bounds $R$ on splitting kernels $P$, $R(q)\ge P(q)$ for all $q$,
algorithm~(\ref{algos:incomplete})
will draw random variables with density
\begin{equation}
{\rm d}T_P(q|Q) = \theta(Q-q) P(q)\Delta_P(q|Q) {\rm d}q\ ,
\end{equation}
where the Sudakov form factor is given by
\begin{equation}
\Delta_P(q|Q) = \exp\left( -\int_q^Q P(t) {\rm d}t  \right) \ .
\end{equation}
\begin{algorithm}
\begin{algorithmic}
\STATE $Q' \gets Q$
\LOOP
\STATE Draw $q$ with density
$$
\theta(Q'-q)R(q)\Delta_R(q|Q'){\rm d}q \ .
$$
\RETURN $q$ with probability $P(q)/R(q)$
\STATE $Q'\gets q$
\ENDLOOP
\end{algorithmic}
\caption{\label{algos:incomplete}The Sudakov veto algorithm as quoted
  in the literature.}
\end{algorithm}
We note here, however, that the algorithm has to be more carefully defined.
Most obviously if, in algorithm~(\ref{algos:incomplete}), 
$P(q)=0$ but $R(q)\ne 0$ for all $q\le q_c$ and some
$q_c$, the algorithm will potentially enter an infinite loop.  We
shall therefore assume that $R(q)$ is suitably restricted to avoid this
situation, making the algorithm well-defined in the sense that it will
never hit a state in which it will not terminate with probability one.

Literally implementing the algorithm as presented above will
not generate the desired density owing to the fact that
${\rm d}T_P$ is not a probability density,
\begin{equation}
\int_q^Q \frac{{\rm d}T_P(t|Q)}{{\rm d}t}{\rm d}t
 = 1- \Delta_P(q|Q) \ne 1 \ .
\end{equation}

At best the algorithm will approximate the target density if, 
for the lowest possible $q$,
$\Delta_P(q|Q)\ll 1$. 
In practice, however, a vanishing $\Delta_P$ will never be encountered in
parton shower simulations, due to the fact that an infrared
cutoff $\mu \ge 0$ is always present. Thus the typically divergent part
of the splitting kernel at $q=0$ is never reached, and the no-emission
probability remains, $\Delta_P(\mu|Q)>0$.

Similarly, the competing processes algorithm
\begin{inlinealgorithm}
\STATE Draw $\{q_i,...,q_n\}$ from ${\rm d}T_{P_i}(q_i|Q)$, $i=1,...,n$
\RETURN $\max (\{q_i,...,q_n\})$
\end{inlinealgorithm}
targeting at drawing random variables with density ${\rm d}T_{P}$,
$P=\sum_i P_i$, will not produce the desired result for the same
reason.

\section{The Complete Algorithm}

\label{sec:Alg}

The failure of the simple algorithm presented in the previous section
has been argued to originate from the fact that the density considered
is not a probability density. 

However, the density
considered in the previous section is also not what is typically aimed at in
a parton shower implementation. (See {\it e.g.} \cite{Gieseke:2004tc} for
a concise treatment.) This can be seen by the fact that
a lower cutoff scale $\mu$ has not been specified, nor is a virtual
no-emission contribution present. Owing to the fact that a parton
shower is to preserve the total inclusive cross section, the combined
density, including both emission and no-emission, has to be a
probability density. As the probability to not emit between two scales
is determined by the Sudakov form factor, the {\it
  probability} density which we are interested in is
\begin{multline}
\label{eqs:sudakov-general}
\frac{{\rm d}S_P(\mu,q|Q)}{{\rm d}q} =
\Delta_P(\mu|Q)\delta(q-\mu)
+\\ \theta(Q-q)\theta(q-\mu)P(q)\Delta_P(q|Q) \ ,
\end{multline}
which relates to the previously introduced density as
\begin{multline}
\frac{{\rm d}S_P(\mu,q|Q)}{{\rm d}q} = \\
\Delta_P(\mu|Q)\delta(q-\mu) + \theta(q-\mu) {\rm d}T_P(q|Q) \ .
\end{multline}
Using sampling by inversion,
\footnote{In this paper {\bf rnd}
denotes a source of uniformly distributed random numbers on
  $[0,1)$.}

\begin{equation}
\label{eqns:inversion}
\int_0^q \frac{{\rm d}S_P(\mu,t|Q)}{{\rm d}t} {\rm d}t = \Delta_P(q|Q)\theta(q-\mu) = {\bf rnd} \ ,
\end{equation}
we find the equation to be solved for the next scale $q$. This is
similar to what one would expect by viewing the Sudakov form factor
$\Delta_P(q|Q)$ as a no-emission probability between two scales $Q$
and $q$, but now explicitly taking into account the dependence on the
infrared cutoff $\mu$.

As the splitting kernel, $P$, is not normally easily integrated,
what is used in actual implementations is instead typically a version of 
the Sudakov veto algorithm where
$\Delta_R(q|Q) = {\bf rnd}$
is solved for some easily integrable function $R(q)>P(q)$, and the radiation
is kept only with a probability of $P(q)/R(q)$. The issue of how to deal with the
fact that the Sudakov factor $\Delta_R(\mu|Q)\ne 0$, however, remains.

In the typically encountered case that $P(q)$ is divergent at an absolute
lower bound (which we take to be $q=0$), the problem with the non-vanishing Sudakov factor 
at the lowest physically considered bound ($q=\mu$) can be circumvented by integrating 
down to $q=0$.
Events which only have emissions
below the lowest physical bound ($\mu$) are then regarded as no-emission events 
\cite{Sjostrand:2006za,Buckley:2011ms}. 
This is guaranteed to work as for such splitting kernels $\Delta_P(0|Q) = 0$.

However, for splittings of massive particles it is the case, that 
- even if the splitting kernel is integrated down to 0 - 
the corresponding Sudakov factor is not vanishing, $\Delta_P(0|Q) \ne 0$. 
This situation can be dealt with by using an overestimation function $R(q)$ which
does correspond to $\Delta_R(0|Q) = 0$. The approximation of a non-divergent
splitting kernel with a divergent one is, however, likely to lead to a 
severe overestimate, i.e. $R(q)\gg P(q)$, which significantly influences
the efficiency of the algorithm.
Alternatively, we here suggest that algorithm (\ref{algos:full}) can be used, 
both for divergent and non-divergent splitting kernels.

\begin{algorithm}
\begin{algorithmic}
\STATE $Q' \gets Q$
\LOOP
\STATE solve {\bf rnd}$=\Delta_R(q|Q')\theta(q-\mu)$ for $q$
\IF{$q=\mu$}
\RETURN $\mu$
\ELSE
\RETURN $q$ with probability $P(q)/R(q)$
\ENDIF
\STATE $Q'\gets q$
\ENDLOOP
\end{algorithmic}
\caption{\label{algos:full}The alternative Sudakov veto algorithm.}
\end{algorithm}

We claim that this algorithm will correctly
produce ${\rm d}S_P(\mu,q|Q)$ for all chosen boundaries $\mu\le
q<Q$. To prove it, we first prove theorem~(\ref{theorems:veto-density}).
\begin{theorem}
\label{theorems:veto-density}
The $q$-density produced by the Sudakov veto
algorithm after $n$ rejection steps and a final acceptance step is given
by
\begin{multline}
\label{eqs:veto-density}
\frac{{\rm d}S_{\text{veto}}^{(n)}(\mu,q|Q)}{{\rm d}q} =
\Delta_R(\mu|Q)\delta(q-\mu)\Delta_{P-R}^{(n)}(\mu|Q)
+\\ \theta(Q-q)\theta(q-\mu)P(q)\Delta_R(q|Q) \Delta_{P-R}^{(n)}(q|Q)
\end{multline}
where
\begin{equation}
\Delta_{P-R}^{(n)}(q|Q) =\frac{1}{n!}\left(-\int_q^Q \left(P(k)-R(k)\right) {\rm d}k\right)^n \ .
\end{equation}
\end{theorem}
From this the correctness of the algorithm follows upon summing over
any number of rejection steps $n=0$ to $\infty$, and the usage of
$\Delta_{R}(q|Q)\Delta_{P-R}(q|Q)=\Delta_P(q|Q)$. Note that
theorem~(\ref{theorems:veto-density}) does include the density of
non-radiating events, and that each time the loop in
algorithm~(\ref{algos:full}) is entered, an event $q$ is drawn from 
${\rm d}S_R$ by virtue of eq.~(\ref{eqns:inversion}).

We will show theorem~(\ref{theorems:veto-density}) using induction
and therefore start by noting that the probability to accept an event,
starting at an intermediate scale $k$, is given by
\begin{multline}
\label{eqs:acceptor}
{\rm d}S^{\text{accept}}(\mu,q|k) = \Delta_R(\mu|k)\delta(q-\mu)
+\\ \theta(k-q)\theta(q-\mu)P(q)\Delta_R(q|k) \ ,
\end{multline}
where the first term reflects the fact that proposal events at the
infrared cutoff are always accepted, while the second term accounts
for proposal events above the cutoff being accepted with probability
$P(q)/R(q)$. For $n=0$ the intermediate scale $k$ equals the
starting scale $Q$, {\it i.e.}
\begin{equation}
{\rm d}S_{\text{veto}}^{(0)}(\mu,q|Q) = {\rm d}S^{\text{accept}}(\mu,q|Q) \ ,
\end{equation}
proving eq.~(\ref{eqs:veto-density}) for $n=0$.  If the algorithm had
performed one rejection step, events could only have been proposed
above the infrared cutoff (otherwise the algorithm would have
terminated), and we have
\begin{multline}
{\rm d}S_{\text{veto}}^{(1)}(\mu,q|Q) =\\
\int_\mu^Q {\rm d}S^{\text{accept}}(\mu,q|k) 
\left(R(k)-P(k)\right) \Delta_R(k|Q) {\rm d}k =\\
\int_\mu^Q {\rm d}S_{\text{veto}}^{(0)}(\mu,q|k) 
\left(R(k)-P(k)\right) \Delta_R(k|Q) {\rm d}k \ .
\end{multline}
Here, the factor of $R(k)-P(k)$ originates from the veto probability,
$1-P(q)/R(q)$, times the kernel $R(q)$ which had been used for the
proposed event. 

To arrive at the desired density in eq.~(\ref{eqs:veto-density})
we use the `chain' property of the Sudakov
form factors,
\begin{equation}
\Delta_R(q|k)\Delta_R(k|Q) = \Delta_R(q|Q) \ ,
\end{equation}
and the relation

\begin{equation}
\Delta^{(1)}_{P-R}(q|Q) = \int_q^Q \left(R(k)-P(k)\right){\rm d}k \ .
\end{equation}
This proves eq.~(\ref{eqs:veto-density}) for $n=1$.
In general,
\begin{multline}
{\rm d}S_{\text{veto}}^{(n+1)}(\mu,q|Q) =\\
\int_\mu^Q {\rm d}S_{\text{veto}}^{(n)}(\mu,q|k) 
\left(R(k)-P(k)\right) \Delta_R(k|Q) {\rm d}k
\end{multline}
reflecting an initially proposed event $k$ below $Q$, which initiated
a sequence of $n$ veto steps and a final acceptance step. Thus, if the
theorem was correct for some $n>0$, we readily obtain the claimed
result for $n+1$ upon using
\begin{multline}
\frac{1}{n!}\int_q^Q \left(\int_q^k f(k') {\rm d}k'\right)^n f(k){\rm d}k = \\
\frac{1}{(n+1)!}\left(\int_q^Q f(k) {\rm d}k\right)^{n+1} \ .
\end{multline}
The competing processes algorithm in turn reads
\begin{inlinealgorithm}
\STATE Draw $\{q_i,...,q_n\}$ from ${\rm d}S_{P_i}(q_i|Q)$, $i=1,...,n$
\RETURN $\max (\{q_i,...,q_n\})$
\end{inlinealgorithm}
which is easily proven as 
${\rm d}S_{P_i}(q_i|Q)$ now is a true probability density.

\section{Towards Splitting Kernels of Indefinite Sign}
\label{sec:Indefinite}

For the remainder of this note we shall be concerned with seeking
solutions to the case of non-positive definite splitting kernels. For
potentially negative-valued `densities' $D(x)$, a Monte Carlo
implementation is still sensible by sampling events $x$ according to
$|D(x)|$ and afterwards assigning weights $+1$ or $-1$, depending on
whether $D(x)>0$ or $D(x)<0$, however, the generalization
of the Sudakov veto algorithm is not obvious.

In this section we will outline an algorithm,
algorithm~(\ref{algos:indefinite}), which is able to deal with the
general case of non-positive definite splitting kernels, but is
limited to considering distributions at fixed starting scale $Q$.
That this is a limitation can be seen from the fact the the generated
density will multiply a $Q$-dependent normalization factor smaller
than one. As long as only one starting scale is considered, this scale
dependence can trivially be normalized away. However, in the case of
varying scales, one would have to introduce scale dependent event
weights larger than one. For the case of an unlimited number of
emissions driven by subsequently sampling the density at varying
scales, there is clearly no upper bound for the combined size of these
weights.  The algorithm presented here could, however, be of practical
interest for cases where splitting kernels P of indefinite sign are
present only for a limited number of emissions.  Such scenarios would
indeed give rise to an upper bound on the expected event weight;
particularly one could consider matrix element corrections
incorporating higher order corrections with the need for appropriate
subtractions to regularize infrared divergences.

To be precise, we
decompose the non-positive definite kernel $P(q)$ as $P^+(q)-P^-(q)$,
where
\begin{equation}
P^\pm(q) = \left\{ \begin{array}{lr} \pm P(q) & :\ P(q) \gtrless 0 \\ 0 & :\ \text{otherwise} \end{array} \right.
\end{equation}
and utilize algorithm~(\ref{algos:indefinite}).
\begin{algorithm}
\begin{algorithmic}
\LOOP
\STATE Draw $q_+$ from ${\rm d}S_{P^+}(\mu,q|Q)$
\STATE Draw $q_-$ from ${\rm d}S_{P^-}(\mu,q|Q)$
\STATE $q\gets\max (q_+,q_-)$
\IF{$q=\mu$}
\RETURN $\mu$ with weight $+1$
\ENDIF
\STATE Draw $t$ from ${\rm d}S_{2 P^-}(\mu,t|q)$
\IF{$t=\mu$}
\IF{$\max (q_+,q_-)=q_+$}
\RETURN $q$ with weight $+1$
\ELSE
\RETURN $q$ with weight $-1$
\ENDIF
\ENDIF
\ENDLOOP
\end{algorithmic}
\caption{\label{algos:indefinite}The algorithm for splitting kernels
  of indefinite sign. See text for the definition of $P_\pm$.}
\end{algorithm}

Note that all random variables needed from Sudakov-type
distributions are readily generated using the veto algorithm as
outlined above. We claim that the algorithm will generate
$$
{\rm d}S_{P}(\mu,q|Q) \times \Delta^2_{P^-}(\mu|Q) \ .
$$
To prove this, note that if $q_+=q_-=\mu$ we obtain a contribution
\begin{multline}\nonumber
\delta(q-\mu)\Delta_{P^+}(\mu|Q)\Delta_{P^-}(\mu|Q) = \\
\delta(q-\mu)\Delta_{P}(\mu|Q)\Delta^2_{P^-}(\mu|Q) \ .
\end{multline}
The probability for $t=\mu$, {\it i.e.} not to re-enter the loop is
clearly given by $\Delta^2_{P^-}(\mu|q)$. Then, if $q_+>q_-$ (and hence $q_+>\mu$),
we find a contribution
\begin{multline}\nonumber
\theta(q-\mu) {\rm d}T_{P^+}(q|Q)\Delta_{P^-}(q|Q)\Delta^2_{P^-}(\mu|q) = \\
\theta(q-\mu) {\rm d}T_{P^+}(q|Q)\Delta_{-P^-}(q|Q)\Delta^2_{P^-}(\mu|Q) \ .
\end{multline}
Finally, if $q_->q_+$ (and hence $q_->\mu$), while including the
negative weight for these events, the last contribution is
\begin{multline}\nonumber
-\theta(q-\mu) {\rm d}T_{P^-}(q|Q)\Delta_{P^+}(q|Q)\Delta^2_{P^-}(\mu|q) = \\
\theta(q-\mu) {\rm d}T_{-P^-}(q|Q)\Delta_{P^+}(q|Q)\Delta^2_{P^-}(\mu|Q) \ ,
\end{multline}
completing the proof.
For the case of several available processes we can always
decompose
\begin{equation}
\label{eqs:pmidecomposition}
P(q) = \sum_i P_i(q) = \sum_i P_i^+(q) - \sum_i P_i^-(q) \ ,
\end{equation}
such that the proposal events $q_\pm$, as well as the `control variate'
$t$ may be generated using the competing processes algorithm for the
individual positive and negative contributions,
\begin{equation}
P^\pm(q) = \sum_i P^\pm_i(q) \ .
\end{equation}

\section{Interleaving Vetoing and Competition}

The algorithm outlined in the previous section may be used to deal
with the case of non-positive definite splitting kernels in full
generality provided we are interested in distributions for a single
starting scale $Q$, or are prepared to accept potentially
large weights. For practical purposes, we are, however,
interested in cascades at subsequent scales $q_1>q_2>...>q_n$, where
$q_{k-1}$ serves as the starting scale of the distribution for
$q_k$. The $Q$-dependent normalization $\Delta_{P-}^2$ present in the
distribution generated will thus make it non-ideal
in the context of cascades.

Here we consider the typically encountered physical
setup for which we may assume that
\begin{equation}
  P(q)=\sum_{i} P_{i}(q)>0 \ ,
  \label{eq:sud}
\end{equation}
still allowing for a probabilistic interpretation, though a subset of
the splitting kernels are of indefinite sign. $P(q)$ can be decomposed
as in eq.~(\ref{eqs:pmidecomposition}), and we can directly identify an
overestimate to the desired splitting kernel,
\begin{equation}
P^+(q) \ge 
P(q) = P^+(q) - P^-(q)  \ .
\end{equation}

This suggests a two-step procedure of interleaving competing processes
and vetoing, formalized in algorithm~(\ref{algos:interleaved}), which we
choose to call the `interleaved veto/competition algorithm'.
\begin{algorithm}
\begin{algorithmic}
\STATE $Q' \gets Q$
\LOOP
\STATE Draw $\{q_i,...,q_n\}$ from ${\rm d}S_{P^+_i}(q_i|Q)$, $i=1,...,n$
\STATE $q\gets \max (\{q_i,...,q_n\})$
\IF{$q=\mu$}
\RETURN $\mu$
\ELSE
\RETURN $q$ with probability $(P^+(q)-P^-(q))/P^+(q)$
\ENDIF
\STATE $Q'\gets q$
\ENDLOOP
\end{algorithmic}
\caption{\label{algos:interleaved}The interleaved veto/competition algorithm.}
\end{algorithm}
Here, the $q_i$ may be generated directly, if the
$P_i^+$ allow to. Alternatively the veto algorithm may be used with
overestimates $R_i^+(q)\ge P_i^+(q)$. The correctness of the complete
algorithm is seen by the fact that the first two instructions in the
loop will guarantee that $q$ is distributed according to ${\rm
  d}S_{P^+}$ by the competing process algorithm. In the following steps, 
the obtained density $P^+(q)$ is corrected to
$P(q)=P^+(q)-P^-(q)$ by virtue of the standard veto algorithm.  Note
that this algorithm will neither require negative weights, or introduce a
$Q$-dependent normalization.

\section{Conclusions and Outlook}

We have given a careful analysis of the main Monte Carlo algorithm
entering current parton shower simulations, the Sudakov veto
algorithm.  Especially, we have discussed in detail the importance of
the no emission probability arising as a consequence of an infrared
cutoff, and suggested an alternative formulation,
algorithm~(\ref{algos:full}), which directly includes the dependence on the
infrared cutoff. This algorithm is argued to be more
efficient in the case of a non-divergent splitting kernel.

We also consider possible extensions to the case of splitting kernels
of indefinite sign.  Such splitting kernels are encountered when
trying to extend parton showers beyond the large $N_c$ limit or beyond
leading order.

First, in algorithm~(\ref{algos:indefinite})
we develop a general algorithm for a splitting kernel of indefinite sign. 
Modulo a normalization dependence on the starting scale of the
algorithm, this case may indeed be dealt with in full generality.
The $Q$-dependent normalization, however, prevents efficient usage in the
context of cascades using an ordered chain of scales. 

For the typically encountered case, in which splitting kernels of 
indefinite sign are present, but the sum over all possible splitting 
kernels stays positive, we give, in algorithm~(\ref{algos:interleaved}),
an algorithm interleaving the competing process algorithm with subsequent 
veto steps.

\section*{Acknowledgments}
We would like to thank Stefan Gieseke, Leif L\"onnblad and Torbj\"orn
Sj\"ostrand for useful discussions.  This work has been supported by
the European Union Marie Curie Research Training Network MCnet under
contract MRTN-CT-2006-035606 and the Helmholtz Alliance ``Physics at
the Terascale''.

\bibliography{sudakov-reloaded}

\end{document}